%% file: root.tex
\documentclass[journal]{IEEEtran}

\usepackage{mathtools}
\usepackage{amsmath}  
\usepackage{amssymb} 
\usepackage{xcolor}
\usepackage{subfigure}

\begin{document}

\title{A Modal-Space Method for Online Power System Steady-State Stability Monitoring}
\author{
\IEEEauthorblockN{Bin Wang, Le Xie, Slava Maslennikov, Xiaochuan Luo,  Qiang Zhang, Mingguo Hong
}

\thanks{Corresponding author: Le Xie}
}
\maketitle

\begin{abstract}
This paper proposes a novel approach to estimate the steady-state angle stability limit (SSASL) by using the nonlinear power system dynamic model in the modal space. Through two linear changes of coordinates and a simplification introduced by the steady-state condition, the nonlinear power system dynamic model is transformed into a number of single-machine-like power systems whose power-angle curves can be derived and used for estimating the SSASL. The proposed approach estimates the SSASL of angles at all machines and all buses without the need for manually specifying the scenario, i.e. setting sink and source areas, and also without the need for solving multiple nonlinear power flows. Case studies on 9-bus and 39-bus power systems demonstrate that the proposed approach is always able to capture the aperiodic instability in an online environment, showing promising performance in the online monitoring of the steady-state angle stability over the traditional power flow-based analysis.
\end{abstract}


\section{Introduction}
	\label{sec:intro}
	\input{intro.tex}

\section{A Brief Review of Selected Stability Problems}
	\label{sec:stab}
	\input{stab.tex}
	
\section{Power System Nonlinear Dynamic Model in Modal Space}
	\label{sec:model}
	\input{model.tex}

\section{A Modal Space Method to Estimate SSASL}
	\label{sec:ssasl}
	\input{ssasl.tex}

\section{Case Studies}
	\label{sec:num}
	\input{num.tex}

\section{Conclusion}
	\label{sec:con}
	\input{con.tex}


\bibliographystyle{ieeetr}
\bibliography{ref}

\end{document}

%% file: intro.tex
The principal cause of the Northeast Blackout in 2003 was a lack of situational awareness due to inadequate reliability tools \cite{2003blackout}. The comprehensive and accurate assessment of reliability, i.e. dynamic security assessment (DSA), requires solving computational-extensive numerical integrations which make DSA viable only for offline planning or long-cycle online applications \cite{savu:2014book}. In practical power system planning and operations, there are usually several thousand contingencies and tens or hundreds of interface limits to be calculated. In addition, considering more diverse power flow patterns brought by a high penetration of intermittent energy resources, future DSA is expected to be executed more frequently with a shorter cycle or even approaching real-time. These trends motivate the research needs for fast stability assessment tools to increase the situational awareness and lower the risk of large-scale blackouts. 

A promising way to address such a challenge is the steady-state stability analysis, which only focuses on the steady-state condition of power systems such that the stability assessment can be significantly simplified. The static stability of the base case (without any contingencies) is a necessary condition of system stability and it has been proved practically useful \cite{savu:2014book}\cite{1998cutsem}, including estimating the stability margin as well as other purposes such as the forward reserve procurement. The basic idea is to identify the steady-state voltage stability limit (SSVSL) in a very speedy fashion by solving static power flow equations such that SSVSL can be calculated and monitored online. The online implementation of this type of approaches will guarantee the system operator's awareness of the situation where a margin becomes dangerously low, allowing preparation and execution of necessary control actions. Such an application seems to be more needed in future grids with a high penetration of intermittent renewable energy where system conditions may change more frequently and significantly.

However, there are three major challenges with existing methods for assessing SSVSL when dealing with future power grid with massive renewables: (i) selection of interfaces and scenarios, i.e. which interface is of concern and how to select the sink and the source \cite{dobson:2016areaangle}\cite{yuan:2018virtualbus}; (ii) need for solving a series of power flows for each interface and for each scenario; and (iii) SSVSL is an optimistic stability analysis since small-signal instabilities may still occur without violating SSVSL (to be shown in case study section of this paper). Even regardless of the computational burden involved in (ii), the selections in (i) are always conducted manually, where people’s experiences and understanding about historic behaviors of the power grid play an important role. Such manual selections can be either difficult or unreliable, especially when considering the power flow pattern and critical interfaces may change more significantly and more frequently with loading and operating conditions for future power grids with a high penetration of intermittent distributed energy resources. To this end, an automated, fast and less-optimistic method is proposed in this paper to estimate the steady-state angle stability limit (SSASL), which captures the Saddle-node bifurcation points along certain stressing directions defined by the associated linearized dynamic system. Although these stressing directions may be unrealistic, it is found by extensive numerical studies that the stability margin defined by the proposed approach along any realistic stressing direction would always become low when approaching the stability boundary. 

The key contributions of this paper are summarized below.

\begin{enumerate}
    \item[1)] We explore the nonlinear modal structure of nonlinear power system dynamic model and propose a direct analysis of SSASL leveraging the nonlinear modal structure. We name it the \emph{modal space method}, or \emph{MS method}. 
    \item[2)] The proposed MS method is free of the selection of interface and scenario, and is always slightly conservative compared with the small-signal analysis (SSA).
    \item[3)] We propose an MS method-based online stability monitoring application, which is shown to be less-optimistic than the traditional power flow based SSVSL analysis.
    \item[4)] We perform extensive numerical studies on the 9-bus and 39-bus power systems, compare with the traditional SSVSL analysis and the standard SSA, and demonstrate the potential of the proposed MS method.
\end{enumerate}


%% file: stab.tex
This section briefly reviews a few selected power system stability problems, i.e. TSA, SSA and static voltage stability analysis (VSA), to point out the scope of this work. The review is neither exhaustive nor comprehensive, and we refer interested readers to reference \cite{kundur:2004stability} for a comprehensive review of power system stability problems.

\subsection{TSA, SSA and static VSA}

TSA of nonlinear dynamical power systems is usually considered to be the most realistic dynamic performance analysis, which is used for benchmarking other simplified stability analyses, e.g. small-signal and voltage stability analyses, and evaluating new control schemes, e.g. damping controls. TSA is often defined upon a set of nonlinear differential-algebraic equations (DAEs), whose accurate assessment by a numerical integration can be very computationally-expensive \cite{khaitan:2016numtsa}. Thus, in practical applications, numerical integration based TSA is used for most utilities, while for some utilities it is used as the final check of very few scenarios screened out by fast but not very accurate analyses, e.g. a direct method \cite{chiang:2015stability}. With today's analytical and computing capability, it is still extremely difficult, if not impossible, to implement online TSA for analyzing all contingencies and all potential changes in loading and generation conditions.

SSA is a common simplification of the stability analysis of nonlinear dynamical systems, which is only valid in a small neighborhood of the equilibrium point, i.e. the steady-state condition, since all nonlinearities with power system DAEs are ignored. To evaluate the small-signal stability of power systems under other steady-state conditions or when subject to possible changes of loading and generations, SSA needs to be re-executed, which is also considered to be computational-expensive for the computing resources in today's control room. 

\begin{figure}[tb!]
	\centerline{\includegraphics[width=0.6\columnwidth]{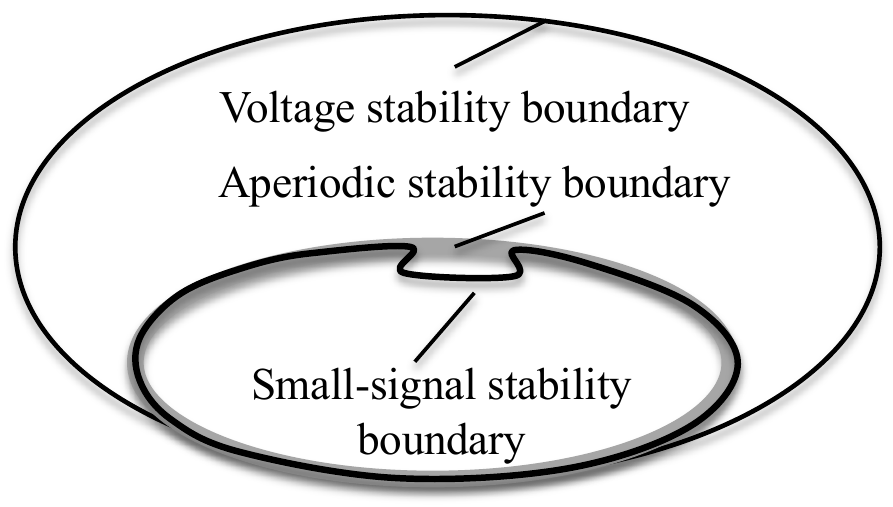}}
	\caption{Boundaries of static voltage stability, small-signal stability and aperiodic stability in parameter space}
	\label{fig:stabboun}
\end{figure}

Static voltage stability analysis (VSA) is a further simplification which ignores all dynamics, while only retaining Kirchhoff's current and voltage laws in the power network, resulting in a set of nonlinear algebraic equations known as power flow equations. Static VSA may imply SSA under some extreme and unrealistic assumptions, while in general there is no such an implication \cite{sauer:1990jacobian}\cite{venikov:1975aperiodic} and voltage stability is usually an optimistic estimate of small-signal stability as shown in Fig. \ref{fig:stabboun}. Most production-level solutions of power flow equations are iterative in nature, e.g. Newton-Raphson and fast decouple methods, although computationally demanding, which are computationally much cheaper than SSA and TSA. 

\subsection{Scope of this work}
\begin{figure}[tb!]
	\centerline{\includegraphics[width=0.6\columnwidth]{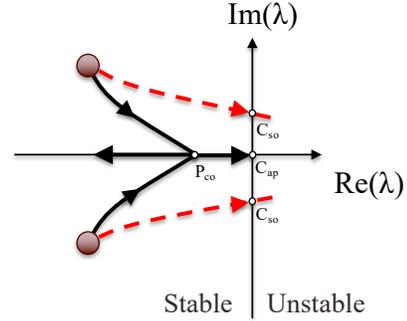}}
	\caption{Two mechanisms of small-signal instability: aperiodic instability (black solid lines) and self-oscillation instability (red dashed lines)}
	\label{fig:ap_so}
\end{figure}

There are two mechanisms for small-signal instability over slow changes of system conditions or parameters \cite{venikov:1975aperiodic}, i.e. \emph{self-oscillation instability} and \emph{aperiodic instability}, which respectively correspond to Hopf-bifurcation and Saddle-node bifurcation in the field of nonlinear dynamics. This work only focuses on the aperiodic instability, i.e. Saddle-node bifurcation induced instability. The red curve in Fig. \ref{fig:ap_so} represents a typical self-oscillation instability, where the real parts of two conjugate eigenvalues change from negative to positive at points $C_{\text{so}}$ heading for an instability without meeting each other. The self-oscillation instability is mostly caused by improper designs of automatic control \cite{venikov:1975aperiodic}, and it is highly related to the damping of oscillatory modes. Therefore, the damping ratio of poorly-damped modes deserve dedicated monitoring programs in operations and planning \cite{zhang:2014phasorpoint} to avoid self-oscillation instability. The aperiodic instability refers to the mechanism that two conjugate eigenvalues first collide on the real axis at point $P_{\text{co}}$, split into two real eigenvalues and one of them becomes positive at point $C_{\text{ap}}$ leading to an aperiodic instability, as illustrated by black curves in Fig. \ref{fig:ap_so}. Note that the aperiodic instability represents a loss of synchronism among generators, i.e. steady-state angle instability \cite{venikov:1975aperiodic}.

This work aims at a novel online steady-state angle stability monitoring application, where the nonlinearity of system dynamics are retained and utilized to directly infer the steady-state angle stability limit (SSASL). The proposed stability analysis is expected to be more accurate, i.e. less optimistic, than static VSA, while being slightly optimistic only when self-oscillation instability occurs before an aperiodic instability, as shown in Fig. \ref{fig:stabboun}.

%% file: model.tex
Consider a classic $N$-machine power system

\begin{equation}  \label{eq:swing}
\left\{  
    \begin{array}{lr}  
             \dot{\delta}_{i}=\omega_{\text{s}}\omega_{i} \\  
             \dot{\omega}_{i}=\frac{1}{2H_{i}}(P_{\text{m}i}-P_{\text{e}i} - D_{i}\omega_{i})  
    \end{array}  
\right.
\end{equation} 

\begin{equation} \label{eq:pe}
\begin{multlined}
P_{\text{e}i} = E_{i}^{2}G_{i} - \sum_{j=1,j\neq i}^{N} \big( C_{ij}\sin(\delta_{i}-\delta_{j}-\delta_{\text{s}ij}) \\
+ D_{ij}\cos(\delta_{i}-\delta_{j}-\delta_{\text{s}ij})\big) 
\end{multlined}
\end{equation}

\noindent where $\delta_{\text{s}ij}$ is the steady-state angle difference between generators $i$ and $j$, $\omega_{\text{s}}$, $P_{\text{m}i}$, $D_{i}$, $H_{i}$, $E_{i}$, $G_{i}$, $C_{ij}$ and $D_{ij}$ are constant, where all loads are represented by constant impedance and included in parameters $C_{ij}$ and $D_{ij}$.

For simplicity, (\ref{eq:swing}) can be re-written as (\ref{eq:fx})

\begin{equation}  \label{eq:fx}
\dot{\mathbf{x}}=\mathbf{f}(\mathbf{x})
\end{equation}

where $\mathbf{x}=\{\delta_{1}$ $\omega_{1}$, $\cdots$, $\delta_{N}$, $\omega_{N}\}$ is the state vector with the equilibrium at the origin and $\mathbf{f}(\mathbf{x})=\{f_{1}(\mathbf{x})$, $f_{2}(\mathbf{x})$, $\cdots$, $f_{2N}(\mathbf{x})\}$ is a smooth vector field.

The models in (\ref{eq:swing}) or (\ref{eq:fx}) are called \emph{the nonlinear dynamic model in angle-speed space} in this paper, while the one to be derived in the following is called \emph{the nonlinear dynamic model in modal space}. Eigen-analysis based on the linearized model of (\ref{eq:fx}) is briefly reviewed below, whose eigenvector matrix will be adopted in deriving the nonlinear dynamic model in modal space. It should be emphasized that although a linear change of coordinate is adopted in our derivation, the nonlinearity of system dynamical model is fully retained without any approximation.

\noindent \textbf{Step 1:} Linearize (\ref{eq:fx}) at the origin and obtain (\ref{eq:Ax}), where $A$ is the Jacobian of $\mathbf{f}$ at the origin.

\noindent \textbf{Step 2:} Calculate the eigenvalues and eigenvectors of $A$ and define the transformation in (\ref{eq:Py}), where $P$ is the matrix consisting of the right eigenvectors of matrix $A$, and $\mathbf{y}$ is the state vector in modal space, i.e. $\mathbf{y}=\{y_{1},y_{2},\cdots,y_{2N}\}$.

\noindent \textbf{Step 3:} Substitute (\ref{eq:Py}) into (\ref{eq:Ax}) such that (\ref{eq:Ax}) becomes (\ref{eq:Ay}), where $\Lambda$ is the matrix consisting the eigenvalues of $A$, i.e. $\Lambda=\text{diag}\{\lambda_{1}, \lambda_{2},\cdots, \lambda_{2N}\}$ and it satisfies $\Lambda=P^{-1}AP$ by definition.

\begin{equation}  \label{eq:Ax}
\dot{\mathbf{x}}=A\mathbf{x}
\end{equation} 

\begin{equation}  \label{eq:Py}
\mathbf{x}=P\mathbf{y}
\end{equation} 

\begin{equation}  \label{eq:Ay}
\dot{\mathbf{y}}=\Lambda\mathbf{y}
\end{equation} 

In a classic $N$-machine power system, the dynamic Jacobian contains $(N-1)$ pairs of conjugate eigenvalues and two real eigenvalues \cite{fabio:2003book}. Without loss of generality, let $\lambda_{2i-1}$ and $\lambda_{2i}$ be a conjugate pair defining the oscillatory mode $i$ for $i = 1, 2,\cdots, N-1$, while $\lambda_{2N-1}$ and $\lambda_{2N}$ be real.

In the eigen-analysis, the linear approximation in (\ref{eq:Ax}) is utilized to study the dynamic behaviors of (\ref{eq:fx}) when subject to small disturbances. As a comparison, the idea to be proposed below will fully maintain all nonlinearities with the original DEs in (\ref{eq:fx}). With the linear change of coordinates in (\ref{eq:Py}), the nonlinear DEs in (\ref{eq:fx}) are transformed to (\ref{eq:gy}), called the dynamic model in modal space. Note that the system in (\ref{eq:gy}) is mathematically equivalent to (\ref{eq:fx}) given the matrix $P$ in (\ref{eq:Py}) is invertible. All subsequent analyses are performed on the nonlinear DEs in (\ref{eq:gy}).

\begin{equation}  \label{eq:gy}
\dot{\mathbf{y}}=P^{-1}\mathbf{f}(P\mathbf{y})\triangleq\mathbf{g}(\mathbf{y})
\end{equation}

\noindent \textbf{Remarks:} 

Corresponding to the concept of the mode in the linear analysis (\ref{eq:Ax})-(\ref{eq:Ay}), the two nonlinear DEs corresponding to a conjugate pair of eigenvalues and their dominated nonlinear dynamic behaviors are referred to as a \emph{nonlinear mode} in this paper. Unlike linear analysis where any two modes and their dynamics are independent, two nonlinear modes are only linearly independent but still coupled nonlinearly, which is termed the nonlinear modal interaction in normal form analysis \cite{vittal:1992nf}, i.e. dynamics initiated in one nonlinear mode may propagate to and affect the dynamics of another nonlinear mode through high-order nonlinear terms. 

It has been proved in \cite{wang:2017motion} that for any classic $N$-machine power system with a uniform damping, the nonlinear dynamics associated with the two real modes do not affect the nonlinear dynamics associated with complex modes. Therefore, the angle stability of systems in (\ref{eq:gy}) are dominated by its first $(2N-2)$ nonlinear DEs of (\ref{eq:gy}). Expanding (\ref{eq:gy}), we can obtain (\ref{eq:gy_exp}) whose first $(2N-2)$ nonlinear DEs in $y_{1},y_{2},\cdots,y_{2N-2}$ are completely decoupled from $y_{2N-1}$ or $y_{2N}$, i.e. the states corresponding to the two real modes.

\begin{equation}  \label{eq:gy_exp}
\left\{  
    \begin{array}{lr}  
             \dot{y}_{1}=g_{1}(y_{1},y_{2},...,y_{2N-3},y_{2N-2}) &  \\  
             \dot{y}_{2}=g_{2}(y_{1},y_{2},...,y_{2N-3},y_{2N-2}) &  \\
             \vdots & \\
             \dot{y}_{2N-3}=g_{2N-3}(y_{1},y_{2},...,y_{2N-3},y_{2N-2}) &  \\
             \dot{y}_{2N-2}=g_{2N-2}(y_{1},y_{2},...,y_{2N-3},y_{2N-2}) &  \\
             \dot{y}_{2N-1}=g_{2N-1}(y_{1},y_{2},...,y_{2N-1},y_{2N}) &  \\  
             \dot{y}_{2N}=g_{2N}(y_{1},y_{2},...,y_{2N-1},y_{2N}) &  \\  
      \end{array}  
\right.
\end{equation} 

%% file: ssasl.tex
In this section, we propose the modal space method to estimate the SSASL of a general classic $N$-machine power system using its nonlinear dynamic model in modal space, as shown in (\ref{eq:gy_exp}). Three versions of the proposed method are developed with different requirements on the system steady-state condition at limit points. The following starts with the introduction of the proposed MS method, continues with a discussion on their differences, and ends up with an introduction of an online stability monitoring application.

\subsection{Simplifying nonlinear dynamic model in modal space}
This subsection presents a simplification of (\ref{eq:gy_exp}) into a number of single-degree-of-freedom (SDOF) nonlinear dynamical systems by using the steady-state condition. Note that the first $(2N-2)$ nonlinear DEs of (\ref{eq:gy_exp}) are still coupled together through nonlinear modal interactions, which place a difficulty for further theoretical analysis of system dynamics when subject to large disturbances. Fortunately, since this work focuses on the steady-state behavior of the system, i.e. how the equilibrium changes nonlinearly with loading condition/generation dispatch, instead of dynamic behaviors \cite{wang:2018nmd}\cite{wang:2015decoupling}, then it is reasonable and helpful to make the following simplification such that (\ref{eq:gy_exp}) becomes (\ref{eq:gy_1mode}): the steady-state condition of the system in (\ref{eq:gy_exp}) implies that when analyzing any nonlinear oscillatory mode $i\in\{1,2,...,N-1\}$, the impact from dynamics of all other modes can be ignored, i.e. $y_{2j-1}=y_{2j}=0$ for all $j\neq i$.

\begin{equation}  \label{eq:gy_1mode}
\left\{  
    \begin{array}{lr}  
             \dot{y}_{2i-1}=g_{2i-1}(0,...,0,y_{2i-1},y_{2i},0,...,0) &  \\
             \dot{y}_{2i}=g_{2i}(0,...,0,y_{2i-1},y_{2i},0,...,0) &  \\
      \end{array}  
\right.
\end{equation}

\subsection{Constructing real-valued DEs for each nonlinear mode}
This subsection applies a linear change of coordinates to (\ref{eq:gy_1mode}) and constructs a real-valued nonlinear SDOF dynamical system for each nonlinear oscillatory mode. The linear change of coordinates is intentionally designed to make the resulting real-valued system similar to a single-machine system, but with a differential characteristic of nonlinearities.

Note that the two state variables and the two nonlinear DEs w.r.t. the nonlinear mode $i$ in (\ref{eq:gy_1mode}) are complex-valued. More specifically, $y_{2j-1}$ and $y_{2j}$ and the two DEs in (\ref{eq:gy_1mode}) are respectively conjugate to each other, which is true if the two eigenvectors corresponding to a conjugate pair of eigenvalues of matrix $P$ in (\ref{eq:Py}) are conjugate to each other. The following adopts the linear change of coordinates \cite{wang:2018nmd} as shown in (\ref{eq:lintrans}) to transform (\ref{eq:gy_1mode}) to a new set of coordinates, such that the two new state variables turn out to be real-valued and possess similar meanings to angle and speed.

\begin{equation}  \label{eq:lintrans}
\left(\begin{array}{c} y_{2i-1} \\ y_{2i} \end{array} \right) = \frac{2}{\lambda_{2i-1}-\lambda_{2i}}  \left(\begin{array}{cc} 1 & -\lambda_{2i} \\ -1 & \lambda_{2i-1} \end{array} \right) \left(\begin{array}{c} \omega_{\text{e}i} \\ \delta_{\text{e}i} \end{array} \right)
\end{equation} 

The determination of the transformation in (\ref{eq:lintrans}) is intuitively generalized from the investigation of a general single-machine-infinite-bus (SMIB) power system \cite{wang:2014freq}\cite{wang:2016fa} as shown in (\ref{eq:smib}), where the system has an equilibrium at origin and $\beta=P_{\text{max}}\omega_{\text{s}}/2H$ for simplicity (note that $P_{\text{m}}=P_{\text{max}}\sin\delta_{\text{s}}$ and $P_{\text{e}}=P_{\text{max}}\sin(\delta+\delta_{\text{s}})$). The linearization of (\ref{eq:smib}) at the origin is obtained as (\ref{eq:smib_lin}). By eigen-analysis, eigenvalues of (\ref{eq:smib})'s Jacobian are $\lambda_{1}=j\sqrt{\beta\cos\delta_{\text{s}}}$ and $\lambda_{1}=-j\sqrt{\beta\cos\delta_{\text{s}}}$, and their right eigenvectors are shown in (\ref{eq:smib_transP}). The linear change of coordinates in (\ref{eq:smib_trans}) can transform (\ref{eq:smib_lin}) to its modal space as shown in (\ref{eq:smib_Ay}), which corresponds to the linearization of (\ref{eq:gy_1mode}). Thus, the transformation in (\ref{eq:lintrans}) is chosen according to the inverse of matrix $P$, as shown in (\ref{eq:smib_pinv}), which can transform (\ref{eq:smib_Ay}) back to (\ref{eq:smib_lin}).

\begin{equation}  \label{eq:smib}
\left\{  
\begin{array}{lr}  
\dot{\omega}= \frac{\omega_{\text{x}}}{2H}(P_{\text{m}} - P_{\text{e}}) = \beta\big(\sin\delta_{\text{s}} - \sin(\delta + \delta_{\text{s}})     \big)\\ 
\dot{\delta}=\omega
\end{array}
\right.
\end{equation}

\begin{equation}  \label{eq:smib_lin}
    \left(\begin{array}{c} \dot{\omega} \\ \dot{\delta} \end{array} \right)
    = \left(\begin{array}{cc} 0 & -\beta\cos\delta_{\text{s}} \\ 1 & 0 \end{array} \right)
    \left(\begin{array}{c} \omega \\ \delta \end{array} \right)
\end{equation} 

\begin{equation}  \label{eq:smib_transP}
    P = \frac{1}{2}\left(\begin{array}{cc} \lambda_{1} & \lambda_{2} \\ 1 & 1 \end{array} \right)
\end{equation} 

\begin{equation}  \label{eq:smib_trans}
    \left(\begin{array}{c} \omega \\ \delta \end{array} \right)
    = P
    \left(\begin{array}{c} y_{1} \\ y_{2} \end{array} \right)
\end{equation} 

\begin{equation}  \label{eq:smib_Ay}
    \left(\begin{array}{c} \dot{y}_{1} \\ \dot{y}_{2} \end{array} \right)
    = \left(\begin{array}{cc} \lambda_{1} & 0 \\ 0 & \lambda_{2} \end{array} \right)
    \left(\begin{array}{c} y_{1} \\ y_{2} \end{array} \right)
\end{equation} 

\begin{equation}  \label{eq:smib_pinv}
    P^{-1} = \frac{2}{\lambda_{1}-\lambda_{2}}\left(\begin{array}{cc} 1 & -\lambda_{2} \\ -1 & \lambda_{1} \end{array} \right)
\end{equation} 

Adopt notations in (\ref{eq:real1}) by separating the real and imaginary parts of (\ref{eq:gy_1mode}) and its eigenvalues, where $a_{i}$, $b_{i}$, $c_{i}$ and $d_{i}$ are real-valued numbers or functions, then substitute (\ref{eq:lintrans}) into (\ref{eq:gy_1mode}) and obtain (\ref{eq:real2}), which is real-valued. Note that the realness of (\ref{eq:real2}) only relies on the conjugate relationships between the two DEs in (\ref{eq:gy_1mode}) and between $\lambda_{2i-1}$ and $\lambda_{2i}$, as shown in (\ref{eq:real1}).

\begin{equation}  \label{eq:real1}
\left\{  
    \begin{array}{lr}  
             \lambda_{2i-1}=a_{i}+jb_{i}  \\
             \lambda_{2i}=a_{i}-jb_{i}  \\
             g_{2i-1}(y_{2i-1},y_{2i})=c_{i}+jd_{i}  \\
             g_{2i}(y_{2i-1},y_{2i})=c_{i}-jd_{i}
      \end{array}  
\right.
\end{equation} 

\begin{equation}  \label{eq:real2}
\left\{  
    \begin{array}{lr}  
             \dot{\omega}_{\text{g}i}=a_{i}c_{i}-b_{i}d_{i} \\  
             \dot{\delta}_{\text{g}i}=c_{i}  
    \end{array}  
\right.
\end{equation}

\subsection{Identifying generalized “power-angle curve” for each nonlinear mode}
It has been found by extensive studies on different system models and under different operating conditions that the equations in (\ref{eq:real2}) always follow a special form as shown in (\ref{eq:real3}), i.e. the derivative of the “generalized speed” is a single-variate function of the “generalized angle” and the derivative of the “generalized angle” is exactly the “generalized speed”. This paper assumes such a relationship without proof. We further define the ``generalized power" by (\ref{eq:Pi}) for the nonlinear mode $i$. Note that the desired information about SSASL is completely maintained in (\ref{eq:Pi}), which will be explored in the next subsection.

\begin{equation}  \label{eq:real3}
\left\{  
    \begin{array}{lr}  
             \dot{\omega}_{\text{g}i}=h_{1i}(\delta_{\text{g}i}) \\  
             \dot{\delta}_{\text{g}i}=\omega_{\text{g}i}  
    \end{array}  
\right.
\end{equation}


\begin{equation}  \label{eq:Pi}
P_{i}=h_{1i}(\delta_{\text{g}i})
\end{equation}

\subsection{Estimating SSASL for each nonlinear mode}
\begin{figure}[tb!]
	\centerline{\includegraphics[width=0.6\columnwidth]{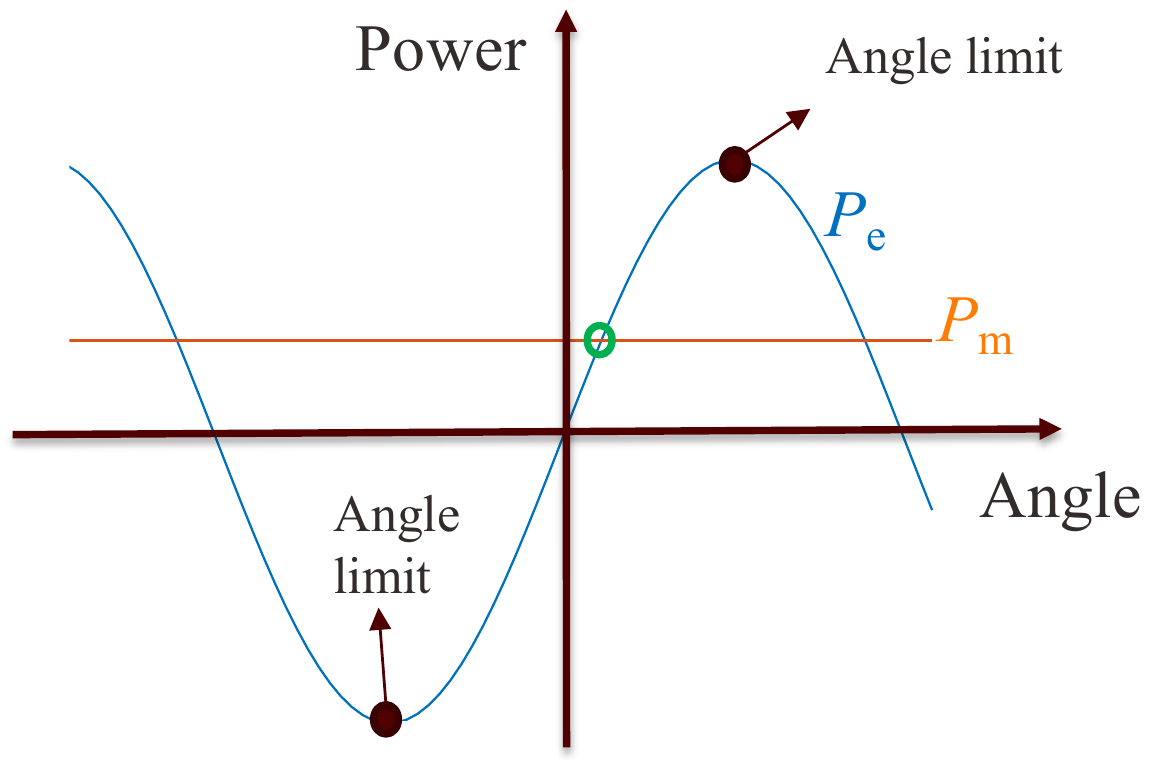}}
	\caption{SSASL points on power-angle curve in an SMIB power system}
	\label{fig:pacurve}
\end{figure}

In an SMIB system, the SSASL is determined by the two closest extreme points on the power-angle curve around the stable equilibrium point as shown in Fig. \ref{fig:pacurve}, which are corresponding to the largest power transfer from (to) the machine to (from) the infinite bus. Similarly, the SSASL of each nonlinear mode can be estimated by identifying the two closest extreme points on the generalized power-angle curve around the origin. Since the generalized power-angle curve of each nonlinear mode is a univariate function in generalized angle as shown in (\ref{eq:Pi}), a bisection based numerical search for these two extreme points, say $\delta_{\text{g}i,1}$ and $\delta_{\text{g}i,2}$, is not very computationally expensive and is therefore adopted in this paper. An approximate analytical solution may be possible and deserves further investigations.

It can be seen that two SSASL points can be obtained for each nonlinear mode. Thus, for the system in (\ref{eq:gy_exp}) having $(N-1)$ nonlinear modes, we have $(2N-2)$ SSASL points in total.

\subsection{Determining system steady state at each limit point}
An SSASL actually represents a system steady state where the system is about to lose stability. This subsection introduces procedures to determine such a system steady state, i.e. voltage magnitudes and angles of all buses, from the estimated SSASL for each nonlinear mode.

Note that the generalized angle and speed, i.e. coordinates of the system in (\ref{eq:real3}), are transformed from original angle and speed derivations, i.e. coordinates of the system in (\ref{eq:swing}), over two linear changes of coordinates in (\ref{eq:Py}) and (\ref{eq:lintrans}). Therefore, each estimated SSASL point represented by the generalized angle and speed can be transformed back to original angle-speed coordinates by the inverse of these two linear transformations, then all other states can be determined accordingly. Denote the two limit points w.r.t. nonlinear mode $i$ as $\delta_{\text{g}i,k}$ with $k=1$ or $2$. Detailed procedures are summarized below. 

\noindent \textbf{Step 1:}  Substitute $(\omega_{\text{g}i},\delta_{\text{g}i})=(0,\delta_{\text{g}i,k})$ into (\ref{eq:lintrans}) and calculate $y_{2i-1}$ and $y_{2i}$.

\noindent \textbf{Step 2:}  Formulate the modal-space state vector as $\mathbf{y}=\{0$, $\cdots$, $0$, $y_{2i-1}$,$y_{2i}$, $0$, $\cdots$,$0\}$.

\noindent \textbf{Step 3:}  Substitute $\mathbf{y}$ into (\ref{eq:Py}) and obtain $\mathbf{x}$.

\noindent \textbf{Step 4:}  Extract all angles from $\mathbf{x}$, obtain $\delta_{1}$, $\delta_{2}$, $\dots$, $\delta_{N}$, and then formulate machine internal voltage vector $\mathbf{E}=\{E_{1}e^{j\delta_{1}}$, $E_{2}e^{j\delta_{2}}$, $\dots$, $E_{N}e^{j\delta_{N}}\}$, where voltage magnitudes are constant which are introduced in (\ref{eq:pe}).

\noindent \textbf{Step 5:} Calculate the terminal current vector $\mathbf{I}_{\text{t}}$ at generator buses by $\mathbf{I}_{\text{t}}=Y_{r}\mathbf{E}$, where $Y_{r}$ is the reduced admittance matrix including generator source impedance and all load impedance.

\noindent \textbf{Step 6:}  Calculate the terminal voltage vector $\mathbf{V}_{\text{t}}$ at generator by (\ref{eq:vt}), where $Z_{\text{s}}=\text{diag}\{Z_{\text{s}1},Z_{\text{s}2},\dots,Z_{\text{s}N}\}$ is the source impedance matrix.

\noindent \textbf{Step 7:} Solve for non-generator bus voltages by (\ref{eq:kcl})-(\ref{eq:vng}).

\noindent \textbf{Step 8:} By now, voltage phasors at all buses have been obtained, as shown in (\ref{eq:vt}) and (\ref{eq:vng}). Other states, including generator power output, load power consumption can be calculated respectively by (\ref{eq:sg})-(\ref{eq:SL}), while line currents and line flows can be calculated accordingly by Ohm's law (omitted here).

\begin{equation}  \label{eq:vt}
\mathbf{V}_{\text{t}}=\mathbf{E}-Z_{\text{s}}\mathbf{I}_{\text{t}}
\end{equation} 

\begin{equation}  \label{eq:kcl}
    \left(\begin{array}{cc} Y_{11} & Y_{12} \\ Y_{21} & Y_{22} \end{array} \right)
    \left(\begin{array}{c} \mathbf{V}_{\text{t}} \\ \mathbf{V}_{\text{non-G}} \end{array} \right)=\left(\begin{array}{c} \mathbf{I}_{\text{t}} \\ 0 \end{array} \right)
\end{equation} 

\begin{equation}  \label{eq:vng}
\mathbf{V}_{\text{non-G}}=-Y_{22}^{-1}Y_{21}\mathbf{V}_{\text{t}}
\end{equation} 

\begin{equation}  \label{eq:sg}
\mathbf{S}_{\text{t}}=\mathbf{V}_{\text{t}}\mathbf{I}_{\text{t}}^{*}
\end{equation}

\begin{equation}  \label{eq:IL}
I_{\text{L}i}=V_{Li}/Z_{\text{L}i}
\end{equation}

\begin{equation}  \label{eq:SL}
S_{\text{L}i}=V_{Li}I_{\text{L}i}^{*}=V_{Li}(V_{Li}/Z_{\text{L}i})^{*}
\end{equation}

We name the above derivation of the system steady state at an estimated limit point as \emph{the first version of the MS method}, or \emph{MS1}, to be distinguished from two more versions named \emph{MS2} and \emph{MS3} to be introduced below. Note that MS implicitly assumes that when the system goes from the current steady state to the steady state at the limit point, (i) generators' internal bus voltage magnitudes maintain unchanged and (ii) loads are modeled as constant impedance. These two implicit assumptions are not very realistic since (i) generators always have automatic voltage regulations to maintain the terminal voltage magnitude to be around a given reference, and (ii) load power consumption is more often used to measure the system stress level, instead of load impedance. Therefore, it would be more reasonable to maintain generator terminal voltage magnitude and load power unchanged. The following will briefly introduce MS2, which guarantees generator terminal voltage magnitude unchanged, and MS3, which guarantees both generator terminal voltage magnitude and load power unchanged.

MS2 starts with steps 1-6 of MS1. Then, the magnitude of each entry of the obtained vector $\mathbf{V}_{\text{t}}$ is overwritten by the corresponding generator terminal bus voltage magnitude under the base-case condition, while all angles of $\mathbf{V}_{\text{t}}$ are intact. The updated generator terminal voltage vector is denoted as $\mathbf{V}_{\text{t}}^{(\text{update})}$. The terminal current is updated by (\ref{eq:itup}) and non-generator bus voltage is updated similarly to (\ref{eq:vng}) but with $\mathbf{V}_{\text{t}}^{(\text{update})}$. Finally, other states can be updated accordingly.

\begin{equation}  \label{eq:itup}
\mathbf{I}_{\text{t}}^{\text{update}}=(Y_{11}-Y_{12}Y_{22}^{-1}Y_{21})\mathbf{V}_{\text{t}}^{\text{update}}
\end{equation} 

MS3 also starts with steps 1-6 of MS1, continues with all additional steps of MS2, then updates the load impedance using the solved bus voltage and the base-case load power consumption (such an operation alters the admittance matrix in (\ref{eq:kcl}), the non-generator bus voltage in (\ref{eq:vng}) and generator terminal current), and finally iterates the updates of generator terminal voltage and current and load impedance until a convergence is reached.

\subsection{Remarks on MS method}
The proposed MS method approach exploits the nonlinearities within the power system nonlinear dynamic model and directly determines SSASL without repetitively solving the nonlinear power flow problem and the linear eigen-analysis. All system states, i.e. a solved power flow condition, at each SSASL point can be determined such that the stability limit can be reflected in all system states, including all transmission line flows and all bus voltages. Therefore, this new method eliminates the need for manually selecting interfaces, since all lines have been considered. In addition, this new approach also eliminates the need for manually selecting scenarios for stressing the system, since each estimated SSASL point represents the largest stressing condition of a nonlinear mode and such stressing directions are uniquely and automatically determined in the proposed method as implied by the generalized power-angle curve (\ref{eq:Pi}). A few more important remarks are summarized below.

\begin{enumerate}
    \item[1)] \textbf{Number:} Two SSASL points can be obtained for each nonlinear oscillatory mode. Therefore, for a classical $N$-machine power system having $(N-1)$ oscillatory modes, there are $(2N-2)$ SSASL points.
    \item[2)] \textbf{Physical meaning:} Each SSASL point is an estimate of Saddle-node bifurcation point, which represents the largest steady-state angle separations without causing steady-state instability when stressing the system about that nonlinear mode, i.e. increasing the power transfer by generation re-dispatch between two groups of generators determined by the mode shape of that mode.
    \item[3)] \textbf{Limitation in handling load change:} As discussed in 2), the proposed approach assumes a constant load model (constant impedance in MS1 and MS2, while constant power in MS3) and then estimates the generation dispatchability limit. However, when the stability limit against load changes is of concern, the proposed approach seems currently not applicable in that regard. One would have to resort to other approaches, e.g. Thevenin's equivalent approach for static voltage stability.
    \item[4)] \textbf{Conservativeness:} It is worth mentioning that the estimated SSASL corresponds to the point $P_{\text{co}}$ in Fig. \ref{fig:ap_so}, instead of $C_{\text{ap}}$. Therefore, along any stressing direction, $P_{\text{co}}$ is always reached before $C_{\text{ap}}$, which makes our proposed method conservative. Numerical studies will show that the MW distance between $P_{\text{co}}$ and $C_{\text{ap}}$ is almost negligible. Thus, the conservativeness of the proposed method is usually small.
    \item[5)] \textbf{Comparison of MS1, MS2 and MS3:} From MS1 to MS3, the computational complexity increases while the accuracy is expected to improve. A brief summary is shown in Table. \ref{table:1}.
    
\end{enumerate}

\begin{table}[!ht]
\centering
\caption{Comparison of MS1, MS2 and MS3} \label{table:1}
	\begin{tabular}{p{0.5cm} |p{2.5cm}| p{2.5cm}| p{1.0cm}}
		\hline
		   & Generator modeling & Load modeling & Iterative? \\ \hline
		MS1& Fixed internal $E_{i}$ & Fixed $Z_{\text{L}i}$  & No  \\ \hline
		MS2& Fixed terminal $V_{\text{t}i}$ & Fixed $Z_{\text{L}i}$ & No \\ \hline
		MS3& Fixed terminal $V_{\text{t}i}$ & Fixed $P_{\text{L}i}$ and $Q_{\text{L}i}$ & Yes  \\ \hline
	\end{tabular}
\end{table}

\subsection{Applying MS method to online stability monitoring}
This subsection presents a way to apply the proposed MS method to the online steady-state angle stability monitoring. For any given operating condition of a classic $N$-machine power system, the proposed MS method can give $(2N-2)$ steady-state conditions respectively representing $(2N-2)$ limit points on the aperiodic stability boundary. Therefore, distances between the current operating condition to those steady-state conditions at limit points can be used as stability margins for the monitoring purpose. Although any states can be used to define the distance, the MW power is usually a practically meaningful measure and is adopted to define the stability margin, as shown in (\ref{eq:Pmargin}).

\begin{equation}  \label{eq:Pmargin}
P_{\text{margin},k}=||\mathbf{P}-\mathbf{P}_{\text{SSASL,}k}||
\end{equation}

\noindent where $\mathbf{P}=\{P_{\text{e}1},P_{\text{e}2},\dots,P_{\text{e}N}\}$ is the generator active power vector at current operating condition, $\mathbf{P}_{\text{SSASL,}k}$ represents the generator active power vector at the condition w.r.t. the $k$-th SSASL point, $k\in \{1,2,\dots,2N-2\}$, $||*||$ is the norm operation ($L^2$-norm is used in this paper).

Thus, the proposed MS method defines $(2N-2)$ MW margins for a given operating condition. If monitoring these MW margins when the system is increasingly stressed, we will have $(2N-2)$ MW margin curves. An instability may occur when any of those curves approaches zero.

%% file: num.tex
This section illustrates the steady-state angle stability monitoring application based on the proposed MS method in detail on the IEEE 9-bus power system \cite{anderson:2003book}, including a comparison to the reference results by static VSA and SSA. The 39-bus power system \cite{39bus} is then tested to show the effectiveness of the proposed method in a larger power system. All computations in this section are conducted in Matlab on a PC with dual-core Intel i5 at 2.6GHz and 8GB memory.

\subsection{Tests on IEEE 9-bus power system}


The IEEE 9-bus power system, whose one-line diagram can be found in \cite{anderson:2003book} and is omitted here, is selected mainly because it is the smallest multi-machine power system whose steady-state angle stability boundary can be visualized in a $2$-D plot, say $P_{\text{e}2}$-$P_{\text{e}3}$ space or $\delta_{21}$-$\delta_{31}$ space. There are three machines, therefore, defining two electromechanical modes. Based on the power flow data from \cite{anderson:2003book}, dynamic data from \cite{wang:2015dae} and using classic generator model, the two electromechanical modes are found to be 1.38Hz and 2.13Hz, where the 1.38Hz mode represents the oscillation between generator 1 and generators 2 and 3, denoted as mode 1, while the 2.13Hz mode is between generator 3 and generators 1 and 2, denoted as mode 2.

The first test is to illustrate the accuracy of limit points identified by the proposed MS method by comparing to reference stability boundaries numerically identified by VSA and SSA. These reference results are obtained by a ray-scanning scheme, whose steps are summarized below for identifying the static voltage stability boundary, while steps for identifying aperiodic (or small-signal) stability boundaries only has a different step 2 where aperiodic (or small-signal) stability is also checked in addition to power flow convergence.

\noindent \textbf{Step 1:} Given an operating condition with a generation dispatch as $\mathbf{P}=\{P_{\text{e}1},P_{\text{e}2},P_{\text{e}3}\}$, select a direction in the $2$-D $P_{\text{e}2}$-$P_{\text{e}3}$ space, say $\mathbf{n}=(\Delta P_{\text{e}2},\Delta P_{\text{e}3})$ and $||\mathbf{n}||=\Delta$, where $\Delta$ represents the step size and $\Delta=10$ MW is used in this paper to initialize the step size for each ray.

\noindent \textbf{Step 2:} Solve power flow with modified generations $P_{\text{e}2}=P_{\text{e}2}+\Delta P_{\text{e}2}$ and $P_{\text{e}3}=P_{\text{e}3}+\Delta P_{\text{e}3}$.

\noindent \textbf{Step 3:} If power flow in step 2 converges, then repeat step 2 to check a farther point along the ray in the $2$-D space. Otherwise, go to step 4.

\noindent \textbf{Step 4:} If $\Delta>\epsilon$\footnote{$\epsilon$ represents the stopping criterion which takes $0.1$MW in this paper.}, then recover the last converged condition by $P_{\text{e}2}=P_{\text{e}2}-\Delta P_{\text{e}2}$ and $P_{\text{e}3}=P_{\text{e}3}-\Delta P_{\text{e}3}$, reduce step size by $\Delta = \Delta/2$ and go to step 2. Otherwise, go to step 5.

\noindent \textbf{Step 5:} Recover generations to the last converged power flow by $P_{\text{e}2}=P_{\text{e}2}-\Delta P_{\text{e}2}$ and $P_{\text{e}3}=P_{\text{e}3}-\Delta P_{\text{e}3}$, and then record $(P_{\text{e}2},P_{\text{e}3})$ as an estimate of the stability boundary.

\noindent \textbf{Step 6:} Go back to step 1 to start over with another direction until all desired directions are searched.

\begin{figure}[tb!]
	\centering    
	\subfigure[$P_{\text{e}2}$-$P_{\text{e}3}$ space]{\label{fig:9_MW_bdry}\includegraphics[width=0.42\columnwidth]{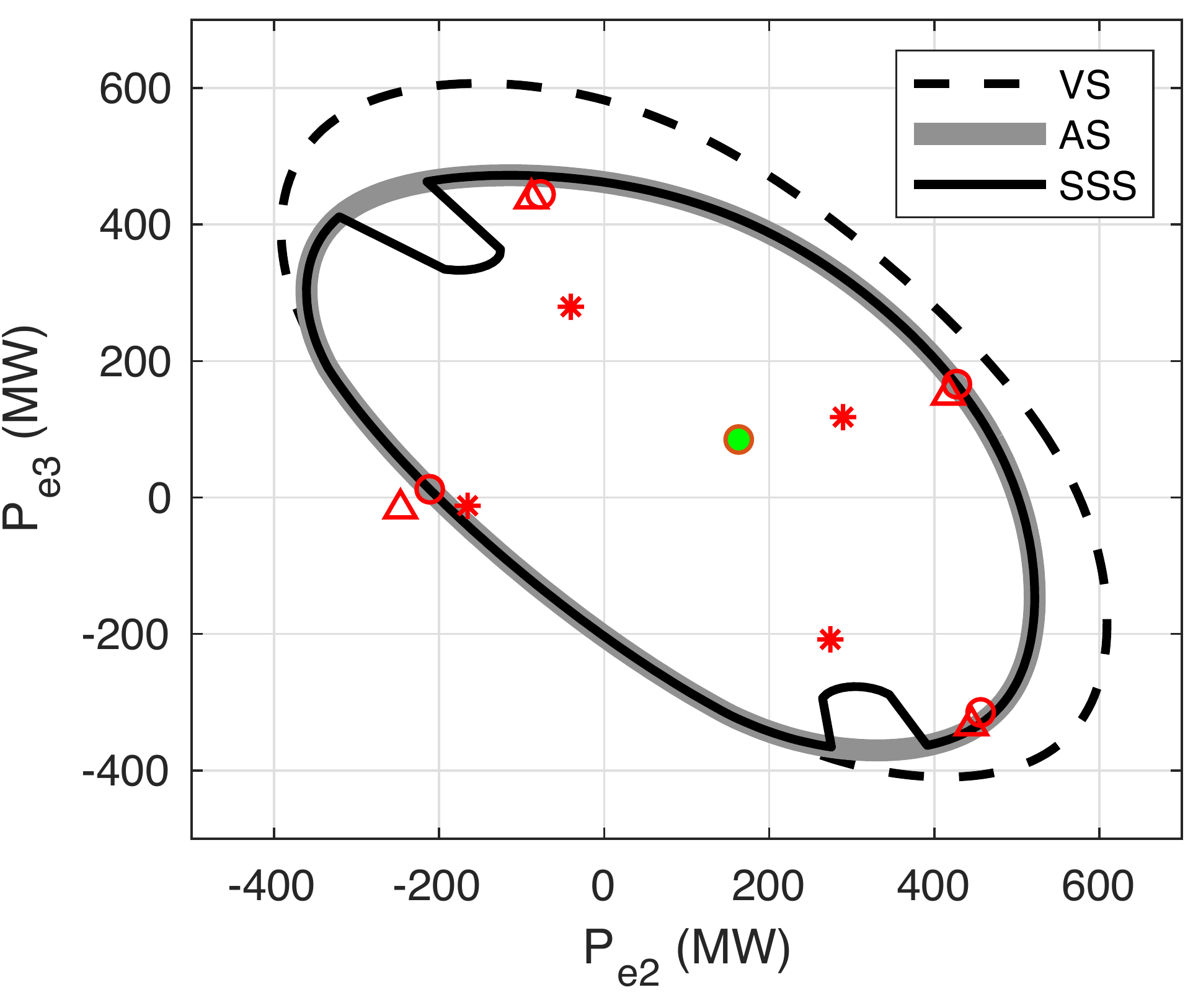}}
	\subfigure[$\delta_{21}$-$\delta_{31}$ space]{\label{fig:9_rad_bdry}\includegraphics[width=0.42\columnwidth]{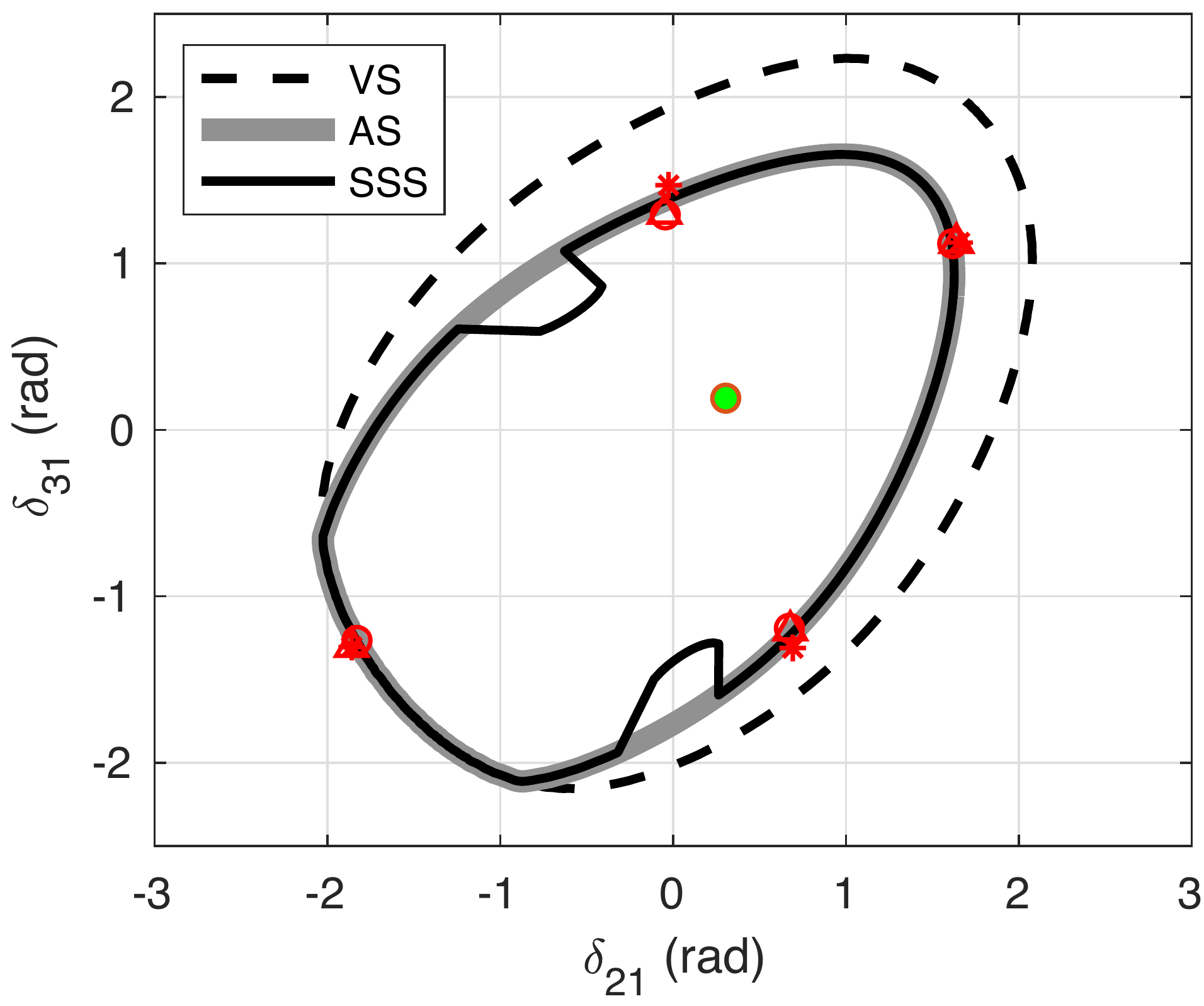}}
	\caption{Numerically identified stability boundaries and SSASL points estimated by MS1, MS2 and MS3 in the 9-bus system. (VS, AS and SSS respectively represent voltage, aperiodic and small-signal stabilities. Red stars, triangles and circles respectively represent the estimated SSASL points by MS1, MS2 and MS 3. The green circle is the given operating condition.)}
	\label{fig:9_bdry}
\end{figure}

Numerically identified reference stability boundaries and the SSASL points estimated by MS1, MS2 and MS3 are shown in Fig. \ref{fig:9_bdry}. Several observations can be obtained: (i) VS boundary is optimistic in general compared to SSS boundary; (ii) AS boundary mostly coincides with SSS boundary while is slightly optimistic when self-oscillation instability occurs before an aperiodic instability; (iii) In $\delta_{21}$-$\delta_{31}$ space, SSASL points estimated by MS1, MS2 and MS3 are fairly close to each other on the AS boundary, showing that all three versions are able to give accurate angle limit; and (iv) In $P_{\text{e}2}$-$P_{\text{e}3}$ space, SSASL points by MS2 and MS3 are fairly close to the AS boundary while those by MS1 are not quite accurate, though being conservative. 

The first test shows how accurate the estimated SSASL points are for a given operating condition by comparing to reference stability boundaries. In fact, if the actual stressing direction does not point to any SSASL points, the system will not exit the stability region through one of these SSASL points estimated at the base-case condition. This is not a problem, since we can always re-calculate these SSASL points by continuously applying the proposed MS method over the change of system steady state. To show an online stability monitoring application based on the proposed MS method, in the second test, the system operating condition is intentionally stressed in a specific direction. A number of steady states, to be checked by the proposed MS method, are selected between the base case and the stability limit.

\begin{figure}[tb!]
	\centering    
	\subfigure[]{\label{fig:9_stre}\includegraphics[width=0.43\columnwidth]{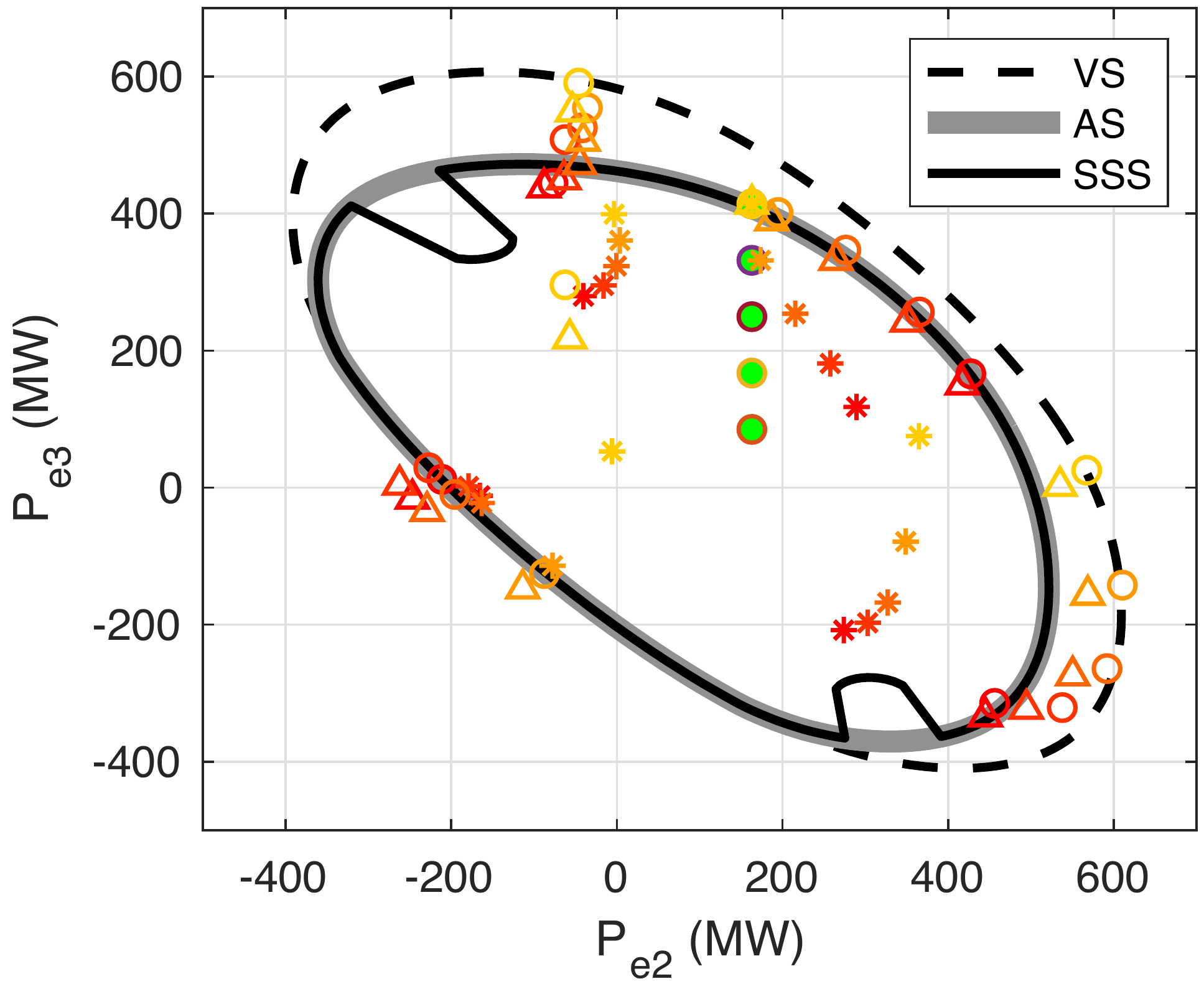}}
	\subfigure[]{\label{fig:9_stre_zm}\includegraphics[width=0.43\columnwidth]{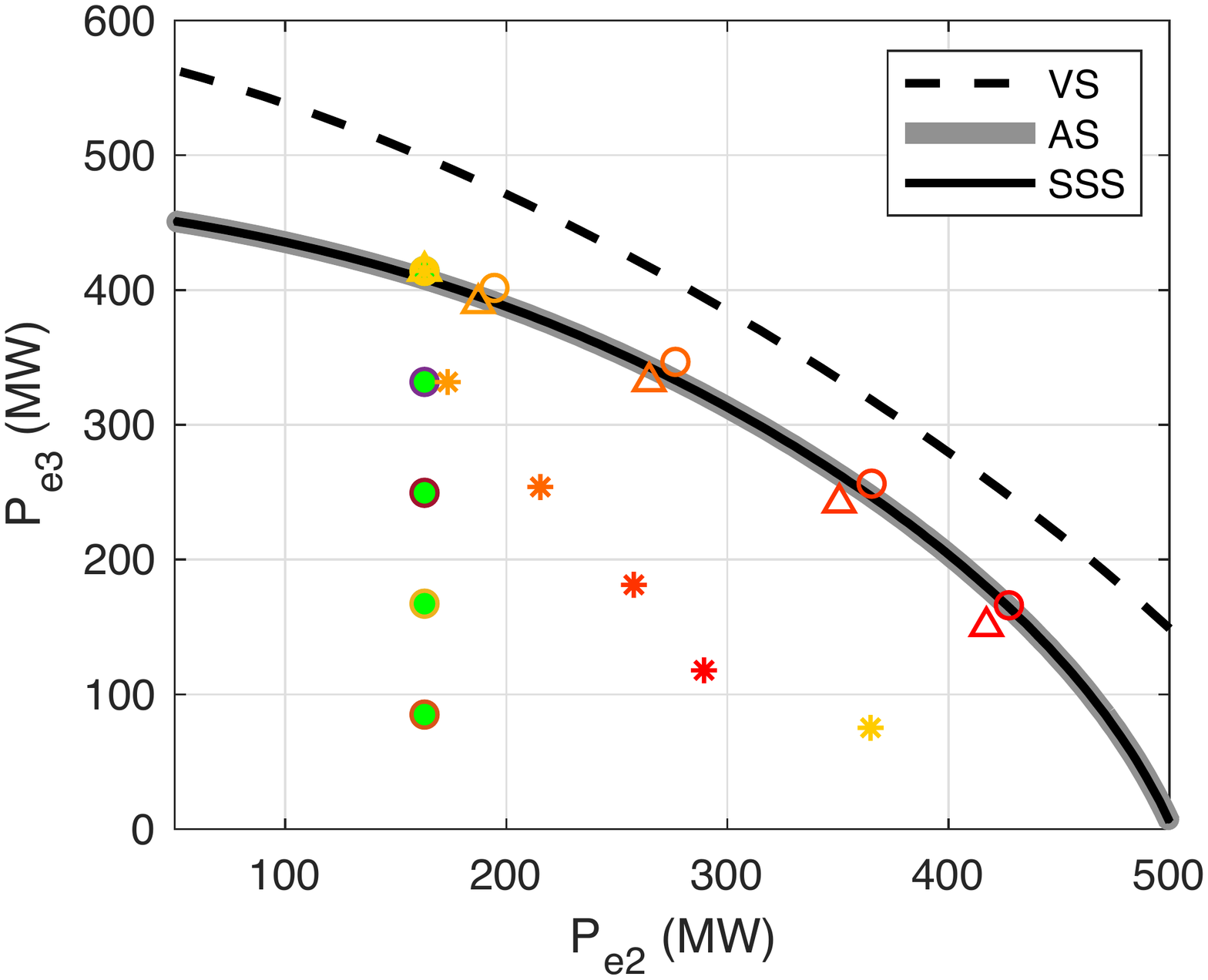}}
	\caption{Estimated SSASL points over stressing.}
	\label{fig:9_stressing}
\end{figure}

Fig. \ref{fig:9_stressing} shows the estimated SSASL points over the stressing process where $P_{\text{e}3}$ increases from $85$MW to $414$MW (all loads and $P_{\text{e}2}$ are maintained unchanged) to cause an instability. Five steady states, from the base case to the limit, are checked by the proposed MS method and the resulting SSASL points change from red to yellow. Fig. \ref{fig:9_stressing} shows that one of the four SSASL points, i.e. the one in the first quadrant, arrests the system when it tries to exit the stability region. It is also observed that when the system gets close to the stability boundary, other SSASL points, than the one arresting the system on the boundary, may not be very accurate. This is fine as long as there is always an SSASL point that accurately arrests the system when it tries to exit the stability region. This has been found true by exhausting all stressing directions in $P_{\text{e}2}$-$P_{\text{e}3}$ space with a small resolution of $2$ degrees.

\begin{figure}[tb!]
	\centerline{\includegraphics[width=0.8\columnwidth]{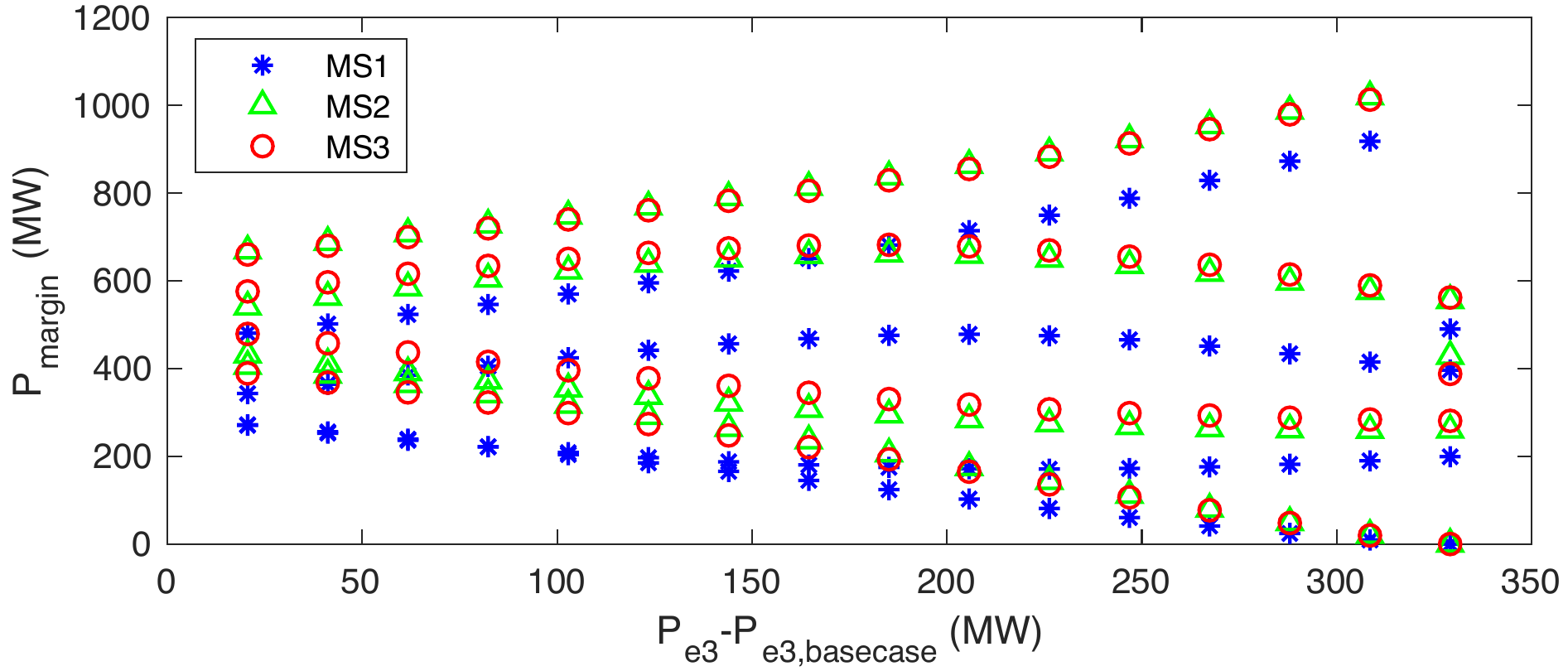}}
	\caption{MW margins v.s. change of $P_{\text{e}3}$}
	\label{fig:9_margin}
\end{figure}

The above visualizes the accuracy of the proposed method and its application in online steady-state angle stability monitoring. However, such a visualization might not be possible for large power systems. To this end, the MW margin defined in (\ref{eq:Pmargin}) can help measure and visualize the distances from an operating condition to SSASL points on the AS boundary. Fig. \ref{fig:9_margin} shows these MW margins when applying the proposed approach to multiple steady states over the stressing process, where the MW margins corresponding to the SSASL point for which the system is approaching are decreasing to zero, while MW margins of other SSASL points are relatively sufficient, either increasing or staying at 200MW or above.

\subsection{Tests on New England 39-bus power system}
\begin{figure}[tb!]
	\centering    
	\subfigure[]{\label{fig:39_case1}\includegraphics[width=0.4\columnwidth]{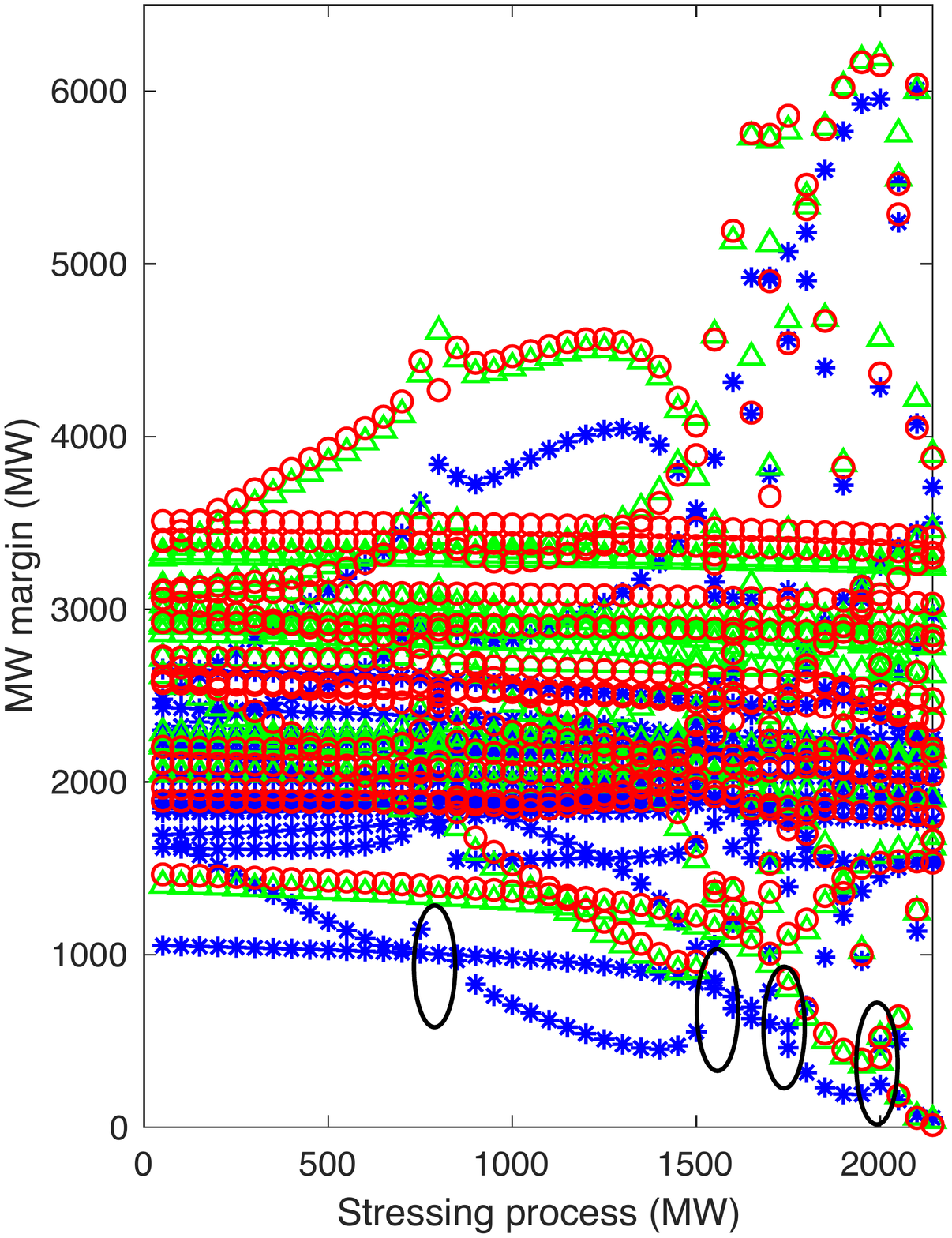}}
	\subfigure[]{\label{fig:39_case1_zm}\includegraphics[width=0.4\columnwidth]{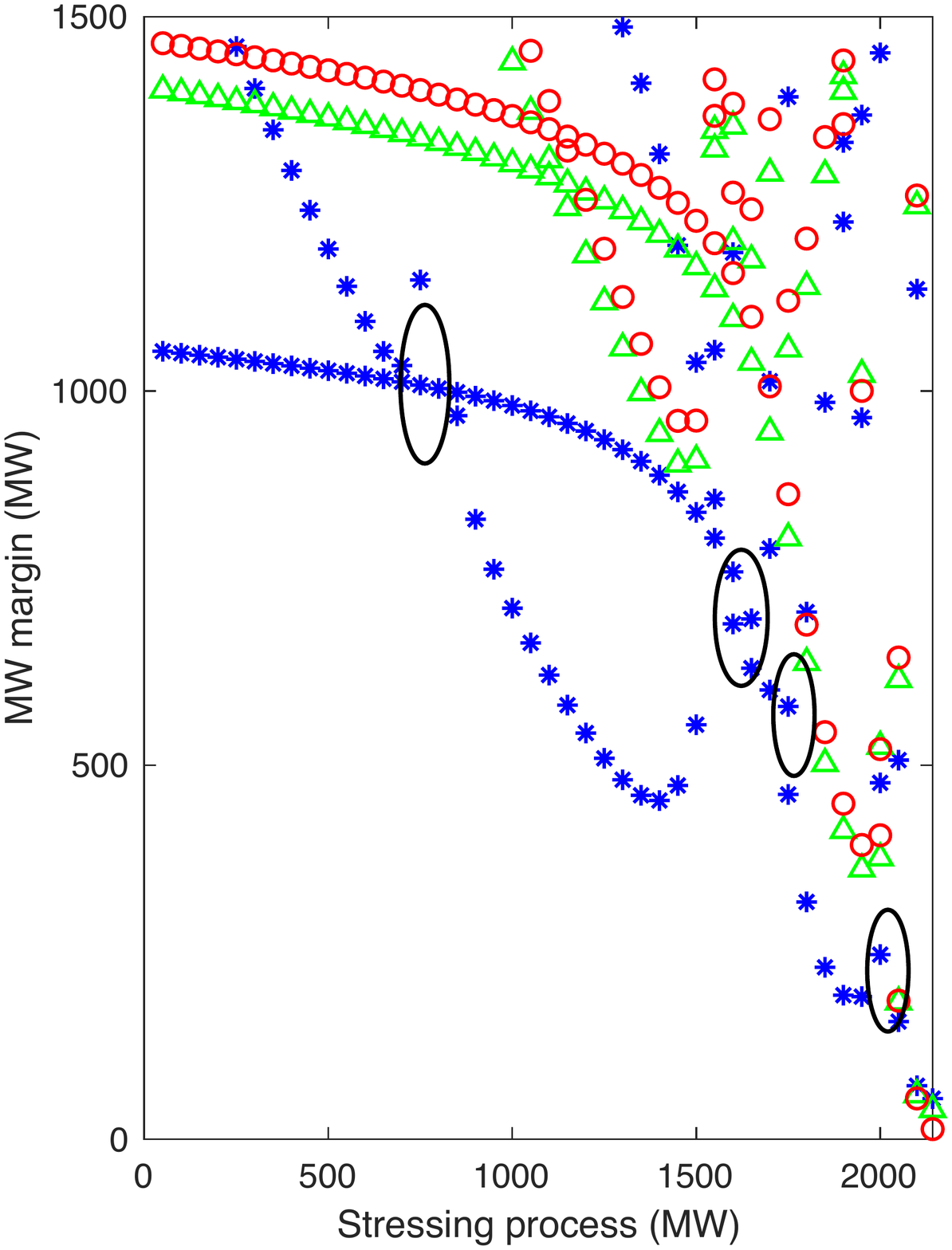}}
	\caption{MW margins in Scenario 1}
	\label{fig:39_1}
\end{figure}

This subsection applies the proposed MS method to a 39-bus power system \cite{39bus}, and two scenarios are tested. In the first scenario, $P_{\text{e}37}$ increases and $P_{\text{e}30}$ decreases by the same MW, resulting in a stress in the local power transfer and causing a small-signal instability when the change of $P_{\text{e}37}$ (or $P_{\text{e}30}$) reaches $2143.04$MW. Along such a stressing direction, the system loses AS and VS respectively when the change of $P_{\text{e}37}$ reaches $2142.95$MW and $2422.42$MW. The MW margins over stressing are monitored by the proposed method and shown in Fig. \ref{fig:39_1}, where MW margins of most SSASL points are above 1000MW even when being close to the instability while only two or three SSASL points may encounter low MW margins over stressing. Note that the smallest MW margin may switch from/to among these two/three SSASL points, as pinpointed in the black circles, over the stressing process, which calls for the need for monitoring all these critical SSASL points to not to miss any potential risk. It is also worth mentioning that (i) aperiodic instability and small-signal instability are extremely close to each other, i.e. less than $0.1$MW in the MW change of the stress, and (ii) right before the system loses it aperiodic stability, the MW margin by (\ref{eq:Pmargin}) using the voltage stability limit is as huge as $482.68$MW, while the MW margins by the proposed MS method are $54.5$MW, $39.79$MW and $13.70$MW respectively for MS1, MS2 and MS3. Similar phenomena can also be seen in scenario 2, where $P_{\text{e}33}$, $P_{\text{e}34}$, $P_{\text{e}35}$, $P_{\text{e}36}$ and $P_{\text{e}38}$ increase and $P_{\text{e}30}$, $P_{\text{e}31}$, $P_{\text{e}32}$ and $P_{\text{e}37}$ decrease, stressing the interface between two distantly located generator groups.

The current implementation of the proposed approach relies on the Symbolic Math Toolbox in Matlab, which turns out to be inefficient. In addition, estimating SSASL points involves the calculation of all eigenvalues and eigenvectors of the linearized system. For a single operating condition of the 39-bus system, it takes up to $30$s. A more efficient implementation without symbolic derivations, along with the selection of critical modes for a partial eigen-analysis are our future work.


\begin{figure}[tb!]
	\centering    
	\subfigure[]{\label{fig:39_case2}\includegraphics[width=0.4\columnwidth]{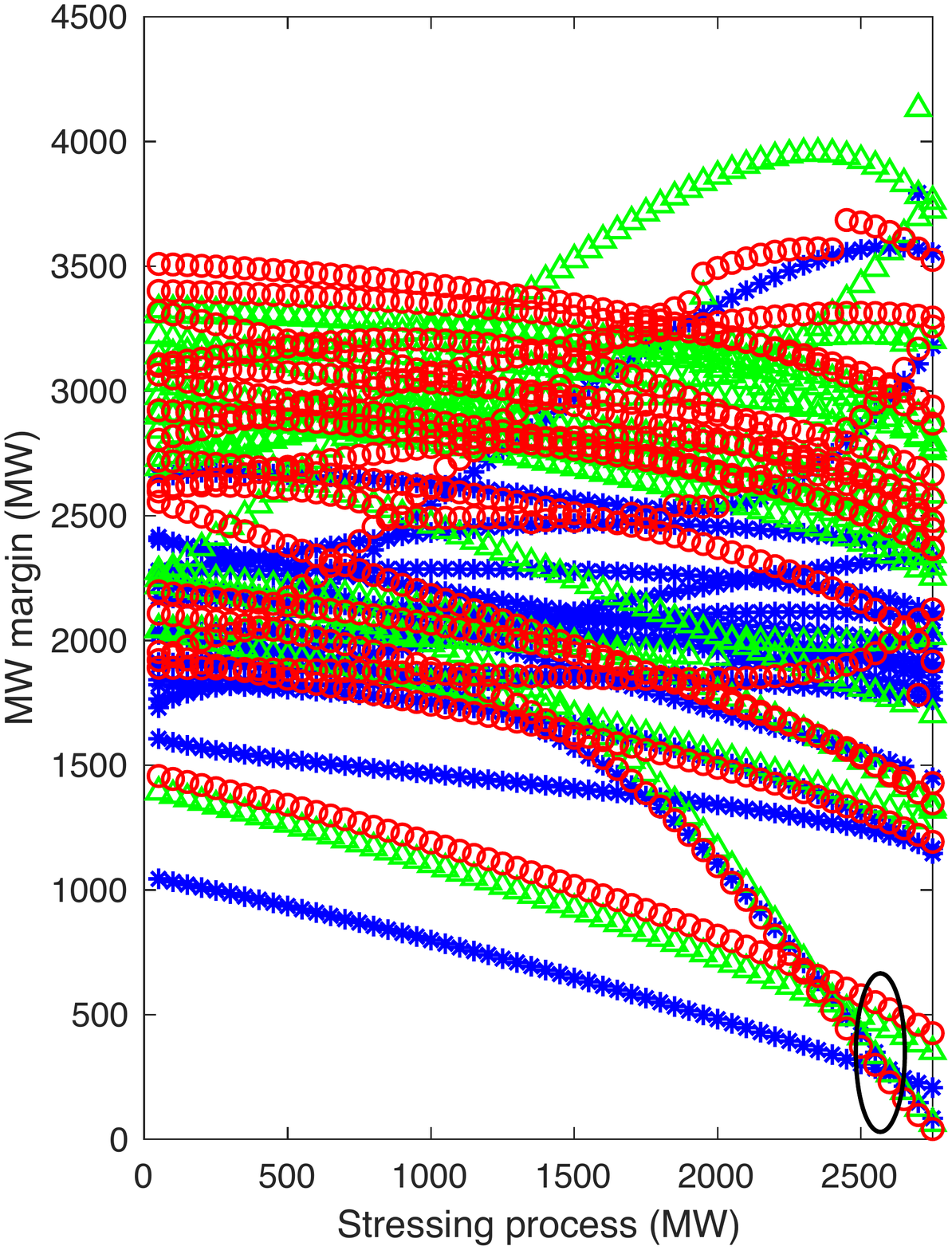}}
	\subfigure[]{\label{fig:39_case2_zm}\includegraphics[width=0.4\columnwidth]{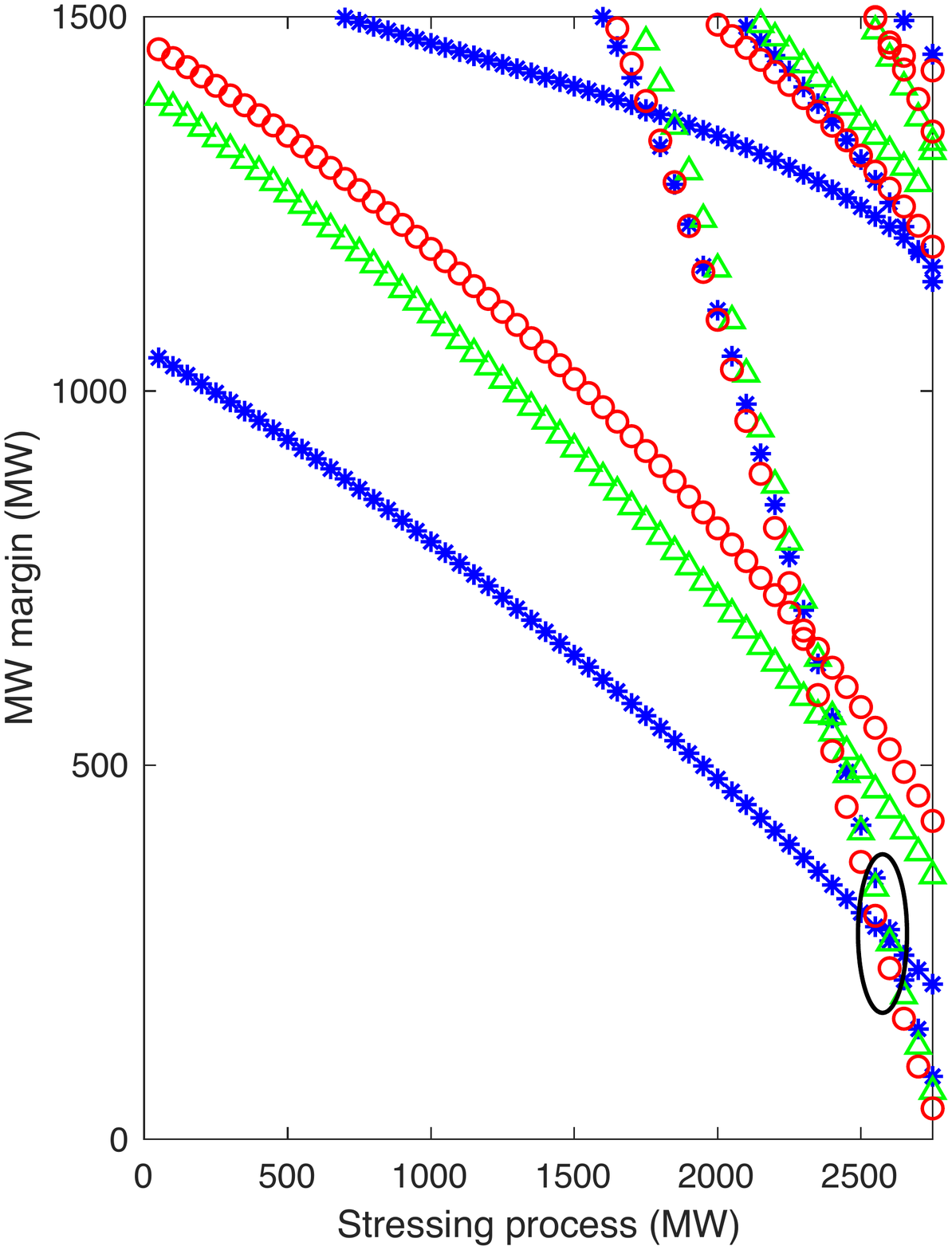}}
	\caption{MW margins in Scenario 2}
	\label{fig:39_2}
\end{figure}

%% file: con.tex
This paper proposes a modal-space (MS) method for estimating the steady-state angle stability limit (SSASL) using the power system nonlinear dynamic model in modal space. The MS method can estimate the SSASL for all system steady states in a single run. A steady-state angle stability online monitoring application is developed based on the MS method and tested on the IEEE 9-bus system and New England 39-bus system. Numerical results show that the proposed MS method is always able to show a meaningful MW margin and arrest the system when it tries to exit the aperiodic stability region.

To further show the potential in online environment, dedicated algorithms without symbolic derivations will be developed to reduce the computation. Future work will also consider $N-1$ contingency analysis and remedial control actions.